\documentclass[runningheads]{llncs}
\usepackage[T1]{fontenc}
\usepackage{graphicx}
\usepackage{booktabs}
\usepackage[misc]{ifsym}

\usepackage{amsmath}
\usepackage{latexsym}
\usepackage{microtype}
\usepackage{booktabs}
\usepackage{adjustbox}
\usepackage{multirow}
\usepackage{IEEEtrantools}
\usepackage{tikz}
\usepackage[misc]{ifsym}

\begin{document}

\title{Guiding Catalogue Enrichment with User Queries}


\newcommand*\samethanks[1][\value{footnote}]{\footnotemark[#1]}
\author{Yupei Du\inst{1}\thanks{Work done while at Amazon}\Letter \and
Jacek Golebiowski\inst{2} \and
Philipp Schmidt\inst{2} \and 
Ziawasch Abedjan\inst{3}\samethanks}

\authorrunning{Y. Du et al.}

\institute{Utrecht University, Utrecht, the Netherlands
\email{y.du@uu.nl}
\and
Amazon, Berlin, Germany 
\email{\{jacekgo,phschmid\}@amazon.com}
\and
BIFOLD \& TU Berlin, Berlin, Germany 
\email{abedjan@tu-berlin.de}}

\tocauthor{Yupei Du, Jacek Golebiowski, Philipp Schmidt, Ziawasch Abedjan}
\toctitle{Guiding Catalogue Enrichment with User Queries}

\maketitle              

\begin{abstract}
    Techniques for knowledge graph (KGs) enrichment have been  
    increasingly crucial for commercial applications that rely on evolving product catalogues.
    However, because of the huge search space of potential enrichment, 
    predictions from KG completion (KGC) methods suffer from low precision, 
    making them unreliable for real-world catalogues. 
    Moreover, candidate facts for enrichment have varied relevance to users. 
    While making correct predictions for incomplete triplets in KGs has been the main focus of  KGC method, 
    the relevance of when to apply such predictions has been neglected. 
    Motivated by the product search use case, 
    we address the angle of generating relevant completion for a catalogue 
    using user search behaviour and the users property association with a product. 
    In this paper, 
    we present our intuition for identifying enrichable data points 
    and use general-purpose KGs to show-case the performance benefits.  
    In particular, we extract entity-predicate pairs from user queries, 
    which are more likely to be correct and relevant, 
    and use these pairs to guide the prediction of KGC methods. 
    We assess our method on two popular encyclopedia KGs, DBPedia and YAGO 4. 
    Our results from both automatic and human evaluations show that 
    query guidance can significantly improve the correctness and relevance of prediction. 
    
\end{abstract}

\section{Introduction}\label{sec:intro}

Knowledge graphs (KGs) have become increasingly prevalent in commercial applications 
to provide accessible and structured representation of knowledge. 
For example, 
shopping websites like Amazon often use KGs to represent product catalogs~\cite{amazon_catalog}, 
where product properties and taxonomies are captured in structures 
similar to the Resource Description Framework (RDF). 
For instance, ``the color of blouse A is red'' 
would be represented as the subject entity ``blouse A'' connecting with the object entity ``red'' 
via the predicate ``color.''
These properties can then be used to offer users recommendations and navigation options 
during product search 
(e.g., ~recommended categories and filtering widgets on the top and the left side 
of the Amazon product search page).

Despite the wide usage of KGs in industry, 
the dynamic nature of commercial applications leads to many practical problems. 
For example, 
sellers may frequently introduce new products without providing the necessary attributes 
or new markets might require different attributions than previous launched markets, 
impeding the maintenance of these KGs.
One approach to remedy this issue is to automatically infer missing information. 
In KG management, this approach is known as KG completion (KGC), 
in which missing information refers to missing triplets. 
Different types of KGC methods have been proposed~\cite{zamini2022review}, 
including methods based on 
mining rules~\cite{galarraga2013amie,galarraga2017predicting,meilicke2019anytime,meilicke2020reinforced}, 
embeddings~\cite{nickel2011three,bordes2013translating,trouillon2016complex,sun2018rotate}, 
and neural networks~\cite{schlichtkrull2018modeling,dettmers2018convolutional,shang2019end}. 
Among these approaches, KG embedding (KGE) methods
have shown good scalability and effectiveness~\cite{chen2020knowledge,wang2021survey}. 

\begin{figure}
    \centering
    \includegraphics[width=0.55\textwidth]{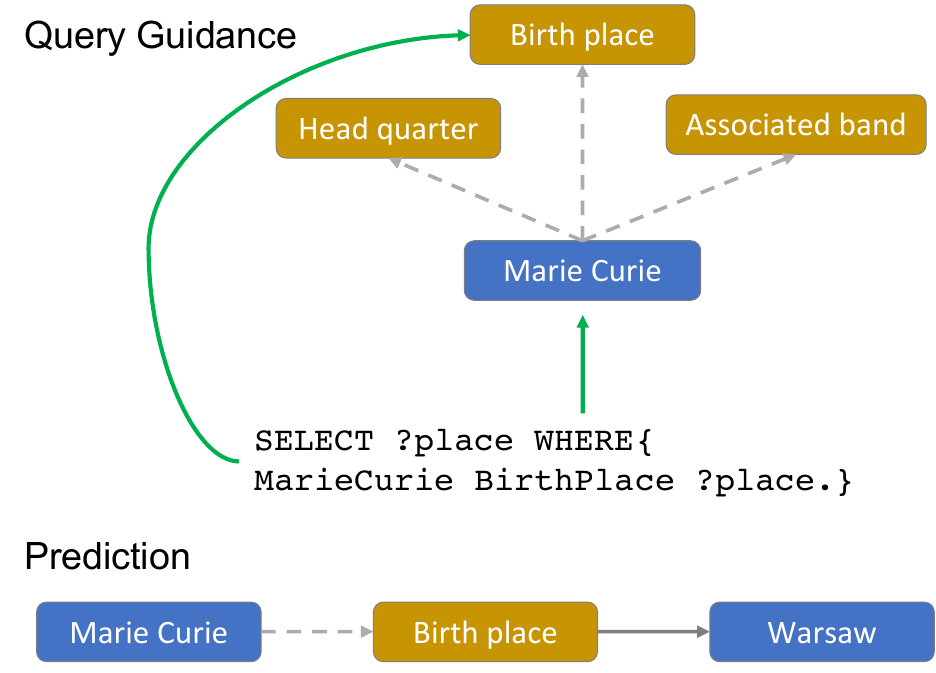}
    \caption{
        An example of using query logs to guide prediction.  
        In this example, we can make prediction on the entity ``Marie Curie" 
        using one of the predicates from ``birthplace", ``head quarter", and ``associated band". 
        Because the query selects the birthplace of Marie Curie, 
        we make predictions from this entity-predicate pair. 
        }
    \label{fig:figure1}
\end{figure}

\paragraph{Limitations in KGC:} Despite the significant advances, 
all of the aforementioned methods suffer from two major issues when predicting missing triplets.  
First, there is a huge space of possible triplets 
when considering all possible combinations of entities and predicates. 
Formally, there are $|\mathcal{E}|\times |\mathcal{P}| \times |\mathcal{E}|$ possible triplets 
just by recombination of existing entities and properties, 
where $|\mathcal{E}|$ and $|\mathcal{P}|$ are respectively the number of entities and predicates in the KG. 
This huge search space makes missing triplet prediction on large-scale KGs 
usually show low precision~\cite{peng2023knowledge}. 
Second, 
KGC methods mostly focus on ensuring the correctness of their predictions, 
by adopting maximum likelihood estimation style objectives. 
However, in real-world scenarios, 
different triplets are usually of different levels of relevance to users. 
With relevance, we refer to the appropriateness and importance of 
a triplet to describe the real-world.
For example, 
although a blouse could have both a size and a manufacturing date, 
knowing its size is more useful for general users than knowing its manufacturing date. 
Therefore, it is also important to take into account which triples are more likely to be used, 
i.e., are more relevant for enrichment. 
\emph{One strategy to mitigate both issues at the same time 
is to provide guidance during the prediction process.} 
For example, 
when already a correct and user-relevant pair of entity-predicates (e.g.~blouse-color) 
are given, one can significantly reduce the search space 
(i.e., from $|\mathcal{E}|\times |\mathcal{P}| \times |\mathcal{E}|$ to $|\mathcal{E}|$). 

In online retail systems, 
it is common to rely on behavioral signals from users to improve their experience. 
For instance, 
we can use users' clicks and purchases to infer their preferences towards various products. 
Regarding the prediction of missing triplets, 
the vast amount of user queries from the product search engine 
can be mined to extract preferences for product attributes. 
For the blouse example, 
we can compare the frequency of queries for 
``<color> blouse" against those for ``blouse released on <date>". 
This data can help make grounded decisions about the relevance and correctness of possible triplets. 

\paragraph{Contributions:}In this paper, 
we propose to guide the missing triplet prediction process 
using user query log signals that express user interests 
to improve the correctness and relevance of the predicted triplets. 
Because commercial KGs and query logs are usually confidential (e.g., ~Amazon product catalogue), 
we show the suitability of our approach on public general-purpose KGs 
using their corresponding SPARQL query logs instead. 
User SPARQL queries usually search for information on general-purpose KGs 
that are relevant to the user and considered correct~\cite{arias2011empirical}.
Thus, they exhibit information  
on how entities should relate to each other and what properties they should display. 
Figure~\ref{fig:figure1} shows a real example from YAGO 4\cite{tanon2020yago}, illustrating that 
the existence of queries can help as a heuristic 
to evaluate the correctness and relevance of properties: 
for the entity ``Marie Curie'', users often query for her ``birthplace'', 
instead of for her ``headquarter'' (incorrect) or for her ``associated band'' 
(taxonomically correct but less relevant for the specific entity ``Marie Curie'', 
because Marie Curie is famous for her scientific rather than musical contributions). 
Although we experiment with general-purpose KGs and query logs in this paper, 
our approach can be easily adapted to commercial KGs. 
Concretely, we make three contributions: 
\begin{itemize}
    \item We propose a simple and efficient method 
    for guiding missing triplet prediction using user queries. 
    We first develop a baseline without user guidance that relies on rejection sampling methods. 
    We then present our query guidance approach for RDF-based KGs (\S\ref{sec:query_guided_triplet_prediction}). 
    Our query guidance method can complement \emph{any} KGC method 
    that make predictions from entity-predicate pairs, 
    which covers most popular KGC methods, including 
    rule-based, KGE , and neural network methods. 
    Our approach can also work with \emph{any} RDF-based KGs, 
    which covers most general encyclopedia KGs and product catalogues. 
    \item We empirically illustrate the benefits of incorporating query guidance. 
    Specifically, we compare our query-guidance method to three baselines: 
    our own baseline that employs rejection sampling without guidance, 
    as well as versions that incorporate two alternative types of guidance, 
    namely KG metadata and KGE scores. 
    This comparison is carried out on two popular general-purpose KGs: 
    DBPedia~\cite{lehmann2015dbpedia} and YAGO 4~\cite{tanon2020yago}, 
    using the popular RotatE KGE model~\cite{sun2018rotate}. 
    Our results from both automatic and human evaluation show that 
    query guidance effectively benefits missing triplet prediction, 
    by selecting entity-predicate pairs that are at least 
    two times more likely to be correct, 
    compared to our baseline without guidance (Table~\ref{tab:human_eval}). 
    \item We build and open-source a dataset consisting of 1600 entity-predicate pairs 
    that are annotated with correctness and relevance scores (\S\ref{sec:results})\footnote{
        Publicly available at \url{https://github.com/LUH-DBS/KGEnrichment}. 
    }.
\end{itemize}

\section{Background and Related Work}\label{sec:preliminaries}

In this section, we describe the relevant studies of KGs, KGE models, 
and rule-based KGC approaches. 
We also introduce RotatE, which is the KGE model used in this paper. 
We further include the notation system used in this paper.

\paragraph{Knowledge Graphs} KGs are structural representations 
of human knowledge in the form of triplets $\mathcal{G} = \{(h, r, t)\}$, 
where $h$, $r$ and $t$ are respectively the subject entity, predicate, and object entity \cite{zamini2022review}. 
For example, ``Marie Curie was born in Warsaw'' will be represented 
as $(\text{``Marie Curie''}, \text{``born in''}, \text{``Warsaw''})$.
Different types of KGs exist, including encyclopedia KGs 
(e.g., DBPedia~\cite{lehmann2015dbpedia} and YAGO 4~\cite{tanon2020yago}), 
domain-specific KGs (e.g., Drugbank~\cite{wishart2006drugbank} and semantic scholar~\cite{semantic_scholar}), 
and task-specific KGs (e.g., Amazon product graph~\cite{amazon_catalog}).   

\paragraph{Knowledge Graph Embeddings and RotatE} 
Various KGE models are proposed in previous studies, 
including translation models~\cite{bordes2013translating,lin2015learning,sun2018rotate}, 
tensor decomposition models~\cite{trouillon2016complex,Kazemi2018SimplE} 
and deep learning models~\cite{wang2014knowledge,jiang-etal-2019-adaptive}. 
These KGE models usually encode entities and predicates in KGs as dense vectors (i.e., ~embeddings), 
which can be used as prior knowledge for downstream tasks 
\cite{sun2019research,zhou2020interactive,sosa2019literature}, 
or to predict missing triplets in KGs~\cite{bordes2013translating,dettmers2018convolutional}. 

In this paper, we focus on RotatE~\cite{sun2018rotate} KGE model.
For each triplet $(h, r, t)$, 
RotatE measures the distance between $h$ and $t$ in the space of $r$ with  
$d_r(\mathbf{h}, \mathbf{t})=\|\mathbf{h} \circ \mathbf{r}-\mathbf{t}\|$, 
where $\mathbf{h}$, $\mathbf{r}$ and $\mathbf{t}$ are the embedding vectors of $h$, $r$ and $t$, 
and $\circ$ is the element-wise product. 
Similar to other KGE models, 
RotatE adopts a margin loss with negative sampling (to facilitate convergence), 
to minimize the distance within existing triplets, 
\begin{IEEEeqnarray}{C}
L=-\log \sigma\left(s_r(\mathbf{h}, \mathbf{t})\right)-
\sum_{i=1}^n \frac{1}{k} \log \sigma\left(-s_r\left(\mathbf{h}_i^{\prime}, \mathbf{t}_i^{\prime}\right)\right), 
\label{eq:rotate_loss}
\end{IEEEeqnarray}
where $\mathbf{h}_i^{\prime}$ and $\mathbf{t}_i^{\prime}$ are the randomly-sampled negative samples, 
$n$ and $k$ are the number and weights of the negative samples respectively, 
and $s_r(\mathbf{h}, \mathbf{t}) = \gamma - d_r(\mathbf{h}, \mathbf{t})$ 
is the margin-based score function with a margin $\gamma$.
One useful attrbute of this objective function is that 
$s_r(\mathbf{h}, \mathbf{t})$ is proportional to $\log{p(\mathbf{h}, \mathbf{t} | \mathbf{r})}$, 
which is illustrated in \cite{mnih2012a} and \cite{yang2020graph}. 

\paragraph{Rule-Based Knowledge Graph Completion}

Besides KGE models, rule-based approaches are also popular in KGC, 
which usually mine compositional rules from statistical cues. 
For example, knowing that a person was born and raised in Amsterdam, 
while also having the knowledge that the official language of Amsterdam is Dutch, 
one can infer that likely this person speaks Dutch. 
Both top-down, which begins from general rule prototypes and specified with data, 
e.g., AMIE~\cite{galarraga2013amie,galarraga2017predicting}, and 
bottom-up, which begins from specific triplets and generalize to rules, 
e.g. AnyBURL~\cite{meilicke2019anytime,meilicke2020reinforced}, are commonly used. 
Compared with KGE models, 
rule-based models are more interpretable but \emph{less scalable}, 
making them hard to apply to large-scale product catalogues.

\section{Query-Guided Triplet Prediction}\label{sec:query_guided_triplet_prediction}

The goal of guided triplet prediction is to increase the utility and accuracy of KGC 
by focusing on triplets that are of interest to users 
and avoiding the generation of potentially irrelevant triplets.
We first introduce a rejection-sampling-based baseline method for predicting missing triplets from KGE models. 
Afterwards, we propose a simple yet effective method 
for guiding the prediction of missing triplets with user query logs 
to obtain triplets of better correctness and relevance. 
In our experiments (\S\ref{sec:results}), 
we apply our query guidance method on the baseline method 
and compare the prediction quality with and without guidance.  

\subsection{Prediction from KGE using Rejection Sampling (\textbf{RS})}

Rejection sampling is a technique for drawing samples from a complex distribution, 
whose unnormalized probability can be expressed as a calculable score $z \cdot p(x)$, 
where $p(x)$ is the probability of a sample $x$, 
and $z$ is a (mostly unknown) normalization factor. 
This method involves using a proposal distribution, denoted as $q(x)$, 
which is easy to sample from (e.g., a uniform distribution). 
Each sample drawn from $q(x)$ is then accepted with a probability 
\begin{IEEEeqnarray}{C}
    p(\text{accept}) = z \cdot p(x) / k \cdot q(x), 
    \label{eq:p_acc}
\end{IEEEeqnarray}
where $k$ is a constant chosen such that $p(\text{accept}) \leq 1$ for all $x$. 
This choice ensures that $p(\text{accept})$ is well-defined. 

Given a trained KGE model, 
because the embedding scores it assigns to triplets are proportional to the probabilities 
(i.e. $s_r(\mathbf{h}, \mathbf{t}) \propto \log{p(\mathbf{h}, \mathbf{t} | \mathbf{r})}$, \cite{mnih2012a,yang2020graph}), 
one can predict missing triplets by sampling from the distribution of KGE scores. 
However, this sampling is not trivial due to that $s_r(\mathbf{h}, \mathbf{t})$ is not normalized. 
One direct fix is to sample entities and predicates uniformly,  
and filter out triplets with low $s_r(\mathbf{h}, \mathbf{t})$ by a pre-defined threshold 
(i.e., regarding triplets with high KGE scores as correct ones). 
However, this threshold can be difficult to determine, 
because the specific relationship between score and triplet quality is unclear. 

As mentioned before, 
rejection sampling can be used to sample from complex distributions, 
as long as unnormalized probabilities are easy to compute, 
making it a good fit for sampling from KGE models. 
Specifically, we can take two steps to predict new triplets by rejection sampling. 
First, according the marginal distribution of predicates in the original KG, 
we sample a predicate $r$. 
Second, for this sampled predicate, 
we sample candidate entities subject $h$ and object $t$ from a uniform proposal distribution. 
We then accept the triplet $(h, r, t)$ with probability 
$p(\text{accept}) = e^{s_r(\mathbf{h}, \mathbf{t})} / e^{\gamma}$, 
where $\gamma$ is the margin from the loss function of RotatE 
(c.f.~\S\ref{sec:preliminaries}). 
The rationales are that, 
1) because $s_r(\mathbf{h}, \mathbf{t})$ is proportional to $\log{p(h, t | r)}$, 
$e^{s_r(\mathbf{h}, \mathbf{t})}$ is an estimation of the unnormalized probability $z \cdot p(h, t | r)$, 
and 2) since $s_r(\mathbf{h}, \mathbf{t}) \leq \gamma$, $e^{\gamma}$ can be seen as 
an unnormalized proposal uniform distribution whose value is greater or equal to $e^{s_r(\mathbf{h}, \mathbf{t})}$ everywhere.\footnote{ 
In practice, 
we can sample a large number of entities pairs and predicates simultaneously, 
and iterate until we accept the specified amount of triplets.}

\subsection{Guided Prediction with Queries (\textbf{QG})}\label{subsec:query_guided_prediction}

Many KGs or catalogues are targets of exploratory search. 
Queries for exploratory search often reflect users' association with an entity, 
e.g., a product and its attributes. 
While a single user might not always hint at the correct signals, 
frequent appearances of certain queries are likely to mirror common expectations of the user base.
Our intuition is to collect such queries and use them to identify gaps in the underlying dataset.

In this paper, we describe our methodology by referring to  SPARQL language, 
which is a popular language for querying RDF data \cite{brickley2004rdf}. 
We make this choice because of the prevalent usage of SPARQL in querying general-purpose KGs, 
including the ones that we experimented with in this study. 

SPARQL supports various functionalities, including 
\texttt{SELECT} (existing triplets), \texttt{CONSTRUCT} (new triplets), \texttt{ASK} (if a triplet exists), 
and \texttt{DESCRIBE} (an entity). 
Among them, 
\texttt{SELECT} queries are similar to actual queries appear in product search. 
\texttt{SELECT} queries usually consist of combinations of predicates and entities, 
where one connecting entity is missing as is queried for. 
For example, the query looking for the birthplace of Marie Curie, 
\texttt{SELECT ?place WHERE\{MarieCurie BirthPlace ?place\}},
would already include Marie Curie as the subject of the triplet 
and birthplace as its predicate. 
Based on our intuition, 
the existence of a query as such suggests that ``Marie Curie'' should have the attribute ``birthplace'', 
which is relevant to users. 
Similarly, product search users usually query for products of certain attributes, 
e.g., \emph{red blouse}. 
This query suggests that all product items from the catalogue of type ``blouse'', 
should have a relevant attribute ``color'', 
knowing that \emph{red} is a type of ``color''
(from named entity recognizers). 
Note that here we adopt pragmatic definitions for ``correctness'' and``relevance'': 
queries show users' interests, and interests imply correctness and relevance 
(we will validate this heuristic in \S\ref{subsec:human_eval}). 
Moreover, we observe that, for both KGs we use in this paper, 
\textbf{more than $95\%$ of the SPARQL queries in the query logs are SELECT queries}. 
We therefore focus on using SELECT queries as guidance. 

Specifically, based on our RS baseline, 
we perform three steps to predict new triplets with the guidance of \texttt{SELECT} queries. 
First, given a \texttt{SELECT} query, we extract all entity-predicate pairs from this query. 
Second, from a uniform proposal entity distribution, 
we sample the second entity for each entity-predicate pair 
Here, we focus on sampling the objects, 
because they are more relevant to the downstream use case of inferring product attributes.
Third, we accept these sampled triplets based on their scores 
computed by the trained KGE model, following Equation~\ref{eq:p_acc}. 
Query guidance therefore help to reduce the prediction space 
from $|\mathcal{E}|\times |\mathcal{P}| \times |\mathcal{E}|$ to $|\mathcal{E}|$, 
which enhances prediction correctness and relevance, 
because they are within the scope of users' interests. 

\paragraph{Comparison with Selecting Top-k Queries}
Another approach of incorporating user query information is 
to select the top-k most frequent queries and make direct predictions from them. 
In contrast to this approach, 
our sampling-based approach additionally considers knowledge from the base KGC method, 
which is a representation of training KG information.
Our approach can be extended for improved performance 
by considering additional aggregation or filtering of user queries. 
However, to illustrate our core idea, the effectiveness of user queries, 
we keep the simplest setting and leave further investigations to future work. 

\section{Evaluation and Results}\label{sec:results}

In this section, 
we evaluate to what extent the guidance of user queries 
can help with missing triplet prediction. 
From the results of both automatic (\S\ref{subsec:automatic_eval}) 
and human (\S\ref{subsec:human_eval}) evaluations, 
we observe that query guidance can dramatically boost 
both the correctness and the relevance of the predicted triplets. 
Moreover, 
to better ground the impact of query guidance, 
we compare query guidance against two alternative types of guidance, 
namely KG metadata (i.e. taxonomy of entities and predicate types) 
and embedding scores from the KGE model (\S\ref{subsec:other_guidance}). 
We observe that, 
although these two types of guidance can both improve prediction quality,
they are outperformed by query guidance.

\subsection{Experimental Setup}\label{sec:exp_setup}

We perform all our experiments using Amazon SageMaker, with a g5.16xlarge instance. 
We use Python 3.7, PyTorch 1.13, DGL 0.4.3, and DGLKE 0.1.2. 
We use RotatE \cite{sun2018rotate} as the KGC model for prediction. 
It took around three GPU days (A10 Tensor Core GPU with 24GB vRAM) 
to perform hyper-parameter optimization of the embedding models 
(20 times of random search on the validation set), 
and four GPU hours to produce all predictions (including the baselines).

\paragraph{KGs and Query Logs}
We use two popular general-purpose RDF KGs for our experiments: 
DBPedia~\cite{lehmann2015dbpedia} English Wikipedia InfoBox 2020.07,\footnote{
        \url{https://databus.dbpedia.org/dbpedia/mappings/mappingbased-objects/2020.07.01}} 
and the YAGO 4 ~\cite{tanon2020yago} English Wikipedia 2020.02.\footnote{
        \url{https://yago-knowledge.org/data/yago4/en/2020-02-24/}}
Moreover, we use 
DBPedia SPARQL Query Logs from March 2021\footnote{
\url{https://devhub.openlinksw.com/pub/Support/44aa7c1b-bd61-4d61-8fef-4075094f62ed/}} 
and YAGO SPARQL Query Logs from 2022\footnote{\url{https://yago-knowledge.org/assets/log\_20221206\_CoQlevVOXUyh.gz}}, 
which were the 
latest ones available at the time of experiments, 
and we removed the queries for entities and predicates that does not exist yet by 2020. 

\paragraph{Pre-processing of KGs and Query logs}
We sanitize the KGs by removing 
entities containing only URLs and numbers, or are lists of other entities 
(e.g. list of all players of a soccer team). 
Beyond conventional pre-processing, we remove all the triplets 
in which both the predicate and at least one entity do not occur in the query logs, 
because these triplets are less relevant to our study.
For example, for the triplet (``Marie Curie", ``birthplace", ``Warsaw"), 
if neither ``birthplace" nor at least one of ``Marie Curie" and ``Warsaw" appear in the query logs, 
we will remove this triplet. 
As a result, we obtain 1.35 million triplets, 881649 entities, and 83 predicates from DBPedia, 
and 12.92 million triples, 4.82 million entities, and 124 predicates from YAGO,  
and will mention them as DBPedia900K and YAGO5M in the remainder of this paper. 
We then randomly split the KGs into train (70\%), dev (10\%) and test (20\%) sets. 

As mentioned before, 
most queries in the logs are SELECT queries ($> 95\%$ for both KGs). 
For example, the SELECT query in Figure~\ref{fig:figure1} 
aims to select the triplets that contain the entity-predicate pair (``Marie Curie", ``birthplace").  
Following the method described in \S\ref{subsec:query_guided_prediction}, 
we extract all entity-predicate pairs from these queries and use them as guidance.
Similar to the pre-processing of KGs,
we only keep the pairs of which 
both the entity and the predicate exist in the processed KGs.
As a result, we obtain 11960 entity-predicate pairs for DBPedia900K,\footnote{
  Wikipedia InfoBox is only a small fraction of the whole DBPedia KG, 
  so most items from the query log is not querying the part of KG that we use. 
} 
and 4.84 million entity-predicate pairs for YAGO5M.

\paragraph{Comparisons}

We primarily show the benefits of query guidance (QG) 
by comparing it with the rejection sampling (RS) baseline. 
We also compare our approach with two alternative types of guidance, 
namely KG metadata and embedding score, 
to better ground the impact of query guidance (details in \S\ref{subsec:other_guidance}). 
\emph{For each method, 
we predict 10 million triplets that are not in the train or dev set.}

\subsection{Automatic Evaluation}\label{subsec:automatic_eval}

\begin{table}
  \centering
  \small
    \caption{
    Automatic evaluation results. 
    \#Hit Triples refers to the number of overlapping triplets 
    between predictions and test sets, 
    and Pair Precision is the precision score 
    of the predicted entity-predicate pairs on test sets: 
    we observe that query guidance (QG) drastically improve the quality of missing triplet prediction, 
    compared with the rejection sampling baseline (RS). 
  }
  \label{tab:precision_results}
  \begin{tabular}{@{}lcccc@{}}
    \toprule
                 & \multicolumn{2}{c}{DBPedia900K} & \multicolumn{2}{c}{YAGO5M} \\
                 & \#Hit Triplets        & Pair Precision       & \#Hit Triplets     & Pair Precision     \\ \midrule
  RS       & 3                & 0.0106               & 0             & 0.0214                      \\
  QG       & \textbf{743}     & \textbf{0.3467}      & \textbf{21}  & \textbf{0.1610}                      \\
  \bottomrule
  \end{tabular}
\end{table}

We first assess the benefits of adopting query guidance by automatic evaluation. 
Specifically, we compute the precision of predictions on the test sets, 
and compare the results for QG against those for the RS baseline. 
We exclude recall scores because 
the same amount of different triplets are predicted for each method 
(i.e. recall is fully dependent of precision). 
In particular, we first evaluate the predicted full triplets. 
Afterwards, we discuss the limitations of evaluating full triplets, 
and include a different setup to evaluate the precision of 
entity-predicate pairs extracted from these predicted triplets. 

\paragraph{Automatic Evaluation of Triplets}

To evaluate the prediction of full triplets, 
we assess the numbers of overlapping triplets (\emph{\#Hit Triplets}), 
i.e., triplets that appear in both predictions and test sets. 
We refrain from using the traditional precision score, because of two reasons.
First, because we predict the same number of triplets for each method, 
the proportions between the number of overlapping triplets is the same as the precision scores. 
Second, as mentioned before, the search space of predictions, 
especially for the RS baseline, is huge 
(e.g., over 60 trillions for DBPedia900K). 
This undesirable attribute can lead to very small precision ratios. 
Considering our relatively small test sets that represent a closed world, 
such numbers might be misleading, 
for being more vulnerable to noises. 

We show the results in the \emph{\#Hit Triplets} columns in Table~\ref{tab:precision_results}, 
and make two observations. 
First, query guidance drastically increases the number of hit triplets, 
i.e., from 3 to 743 on DBPedia900K and from 0 to 21 on YAGO5M. 
The most likely reason for such huge improvements 
is the vast reduction of the search space size, 
from $|\mathcal{E}|\times |\mathcal{P}| \times |\mathcal{E}|$ to $|\mathcal{E}|$, 
where $|\mathcal{E}|$ and $|\mathcal{P}|$ are 
respectively the number of entities and predicates in KG: 
concretely, the search spaces of possible triplets decrease for 
6.77 million and 597.68 million times for DBPedia900K and YAGO5M. 
We note that the larger numbers of overlapping triplets on DBPedia900K, 
compared with YAGO5M,  the larger KG, 
may originate from the same reason: 
the search space of YAGO5M is approximately 137 times larger than DBPedia900K. 
Second, both methods have rather low numbers of overlapping triplets 
(at most hundreds compared to 10 million predicted triplets). 
This observation is consistent with our intuition that 
KGE models usually cannot make accurate predictions on large KGs, 
highlighting the importance of using query guidance.

\paragraph{Automatic Evaluation of Entity-Predicate Pairs}

The evaluation of full triplets has two major limitations. 
First, the quality of predicted full triplets of QG depends on two factors: 
the quality of entity-predicate pairs from user queries, 
and the performance of the KGE model in predicting the second entities. 
It is thus difficult to isolate the benefits of using entity-predicate pairs as guidance. 
Second, as observed in the previous experiment, 
the numbers of overlapping full triplets between predictions and test set 
can be very low for large-scale KGs, 
which makes such comparisons vulnerable to noises. 
For example, our KGE model produces 0 and 21 overlapping triplets 
on YAGO5M using RS and QG respectively: 
it is difficult to understand to which extent QG actually improves the prediction accuracy, 
because such proportions can be susceptible to randomness.
Moreover, predicting entity-predicate pairs themselves is meaningful 
for improving user experience of product search as well: 
shopping websites can notify vendors which missing product attributes are relevant to users, 
so that such information can be manually added, 
which can then be used for search navigation and recommendation.

To accommodate the previous considerations,
we focus on comparing the entity-predicate pairs extracted directly from user queries 
against the ones extracted from the predicted triplets of other methods. 
Specifically, 
we extract all entity-predicate pairs from the predictions of each method, 
and calculate the precision scores on the test sets, 
i.e., the percentage of entity-predicate pairs that consist of at least one triplet from the test sets. 
We show the results in the \emph{Pair Precision} columns in Table~\ref{tab:precision_results}. 
Consistent with our observation on the full triplets, 
the guidance of user queries significantly boosts the prediction accuracy, 
by at least $\sim 8$ times. 
Besides, we observe that the gap between QG and RS is smaller 
compared with the results from the evaluation of full triplets. 
This observation implies that KGE models predict the second entities more accurately 
based on entity-predicates extracted from user queries, 
compared with based on the ones that are randomly sampled. 
In other words, \emph{QG not only offers more correct and relevant entity-predicate pairs, 
but also helps KGE models predict better}.

\begin{table}
\centering
\small
\caption{Human evaluation results. 
    Correct and relevant columns show the precision scores 
    of predicted entity-predicate pairs regarding correctness and relevance. 
    R/C shows the percentage of relevant triplets in all correct triplets. 
    We observe that 
    1) consistent with automatic evaluation, 
    query guidance greatly improves prediction quality over the RS baseline; 
    2) guidance of both KG metadata (KM) and embedding score from KGE models (ES) 
    are beneficial, but outperformed by query guidance; and 
    3) query guidance can also improve the relevant ratio among correct triplets.
    }
\label{tab:human_eval}
\begin{tabular}{@{}lcccccc@{}}
\toprule
       & \multicolumn{3}{c}{DBPedia900K}                       & \multicolumn{3}{c}{YAGO5M}     \\ 
       & Correct         & Relevant       & R/C                & Correct        & Relevant       & R/C              \\ \midrule
RS     & 0.345           & 0.220          & 63.8\%             & 0.305          & 0.225          & 73.8\%           \\
QG     & \textbf{0.950}  & \textbf{0.895} & \textbf{94.2\%}    & \textbf{0.990} & \textbf{0.850} & \textbf{85.9}\%  \\
ES     & 0.425           & 0.355          & 83.5\%             & 0.345          & 0.235          & 68.1\%           \\
KM     & 0.750           & 0.685          & 91.3\%             & 0.920          & 0.760          & 82.6\%           \\ \bottomrule
\end{tabular}
\end{table}

\subsection{Human Evaluation}\label{subsec:human_eval}

Automatic evaluation has the drawback of closed-world assumption: 
because KGs are not complete, 
the triplets in the test sets are only a small fraction of all possible (missing) triplets. 
In other words, 
entity-predicate pairs that are absent from the test set can still be correct and relevant. 
To address this issue, we also conduct a human evaluation of entity-predicate pairs. 
In particular, for each method on each KG,  
we randomly select 200 entity-predicate pairs, 
manually annotate their correctness and relevance, 
and compute the precision scores. 

We use the following general guidelines for annotation: 
1) In \textbf{correct} entity-predicate pairs, 
the entities should be able to logically possess the attribute or relationship described by the predicate. 
An incorrect counterexample is (``saw rock" - ``birthplace"), 
because saw rock, which is a rock in South Atlantic Ocean, is an inanimate object; 
and inanimate objects cannot have attributes like ``birthplace". 
2) In \textbf{relevant} entity-predicate pairs, 
the predicates should provide pertinent information about the entity 
in the context of the knowledge domain that the entities belong to. 
In other words, annotators should evaluate whether 
an average user querying the KG would 
find the predicate's information beneficial or essential to their query purpose. 
For example, (``Starsailor" - ``band member") is a relevant entity-predicate pair, 
because ``Starsailor" is a rock band, 
and users would likely want to know the members of a band they're looking up.
In contrast, (``William Bayliss" - ``associated band") should be annotated as correct but irrelevant, 
because although ``William Bayliss", as a person, can associate with a band, 
he is known for his physiology contributions, not his musical affiliations.

We show our human evaluation results in Table~\ref{tab:human_eval} 
(the rows for RS and QG). 
Besides the precision scores of predicted entity-predicate pairs 
regarding both correctness and relevance, 
we also include the percentage of pairs that are annotated as relevant 
in all pairs that are annotated as correct (R/C).  
We make two observations. 
First, consistent from our observations in automatic evaluations, we observe that 
query guidance can improve both correctness and relevance of the predictions by a large margin
(i.e. from $<0.35$ to $\geq 0.95$ for correctness, and $<0.25$ to $\geq 0.85$ for relevance). 
Notably, besides the absolute numbers of correct and relevant entity-predicate pairs, 
QG also achieves better R/C, indicating that 
\emph{query guidance is beneficial for the relevance of predictions, 
beyond merely enhancing the fraction of correct predictions}.
Second, compared with the precision scores from our automatic evaluation (Table~\ref{tab:precision_results}), 
we observe much higher values in human evaluation. 
We believe that this result validates our previous analyses 
on the closed-world issue of automatic evaluation: 
because KGs are not complete and 
many correct and relevant predicted entity-predicate pairs are not included in the test set, 
precision scores from automatic evaluation are actually lower estimations than reality.

\subsection{Comparison with Other Types of Guidance}\label{subsec:other_guidance}

Beside user queries, there exist other types of information that 
can help identify helpful entity-predicate pairs to guide the missing triplets prediction. 
To better ground the impact of query guidance, 
we compare our approach to two alternative types of guiding information, 
namely KG metadata (KM) and embedding score (ES). 
Consistent with our previous experiments, 
we assess them using both automatic and human evaluations. 
Our results show that, 
although both types of guidance can improve prediction correctness and relevance, 
they are outperformed by QG, 
highlighting the relative advantage of using query guidance.

\paragraph{KG Metadata Guidance (\textbf{KM})}

Both KGs used in this paper provide metadata used to construct them. 
Concretely, they retain the type of each entity, 
and the domain and range of each predicate, i.e., 
which types of entities that the predicate can accept as its subject and object. 
The combination of these two types of metadata can 
help filter out incompatible entity-predicate pairs.
For example, knowing the metadata that 
the predicate ``largest city'' can only accept the entity type ``place'' as subject 
can help us filter out the pair (``Marie Curie", ``largest city"),  
because ``Marie Curie'' is of type ``person'' not ``place''. 
We therefore divide entity-predicate pairs extracted from 
the predicted triplets of the RS baseline into (KG metadata) compatible and incompatible groups, 
and then compute the precision score of each group on the test sets. 

It is worth noting that, 
likewise the incompleteness of the KGs themselves as we have discussed, 
KG metadata can also be incomplete. 
In this case, ``incompatible" entity-predicate pairs can still be correct or relevant. 
For example, for a entity-predicate pair (``Germany", ``largest city"), 
if we only know ``Germany" is a ``country", 
and we do not have the metadata that ``country" is always ``place", 
we will categorize this pair as incompatible.

\paragraph{Embedding Score Guidance (\textbf{ES})}
We also investigate whether embedding scores computed by KGE models 
(i.e. $s$ in Equation~\ref{eq:rotate_loss}) 
can help us select correct and relevant entity-predicate pairs. 
Different from KM, 
which directly divide entity-predicate pairs into two separate groups 
(i.e., compatible and incompatible), 
embedding scores are continuous values. 
For clearer evaluation, 
we divide all entity-predicate pairs predicted by the RS baseline into 50 bins, 
based on the highest embedding score from the triplets that include each pair. 
For instance, if an entity-predicate pair appears in 10 different predicted triplets, 
we use the triplet with the highest score to determine the bin for that pair. 
Similar to KM, we then compute the precision score of each group on the test sets. 

\paragraph{Usage of KM and ES}
In contrast to the query guidance approach, 
neither KM nor ES directly offer entity-predicate pairs for KGE models to make predictions on.  
Instead, they provide guidance in a post-hoc way, 
by either judging whether a predicted entity-predicate pair is compatible with KG metadata (KM), 
or assigning this pair a continuous embedding score 
(i.e. $s$ in Equation~\ref{eq:rotate_loss}), 
whose quantity indicates how likely this pair is correct (ES). 
Therefore, we apply them on the 10 million triplets predicted by the RS baseline as filters 
to select more possible triplets and entity-predicate pairs. 

\begin{table}
  \small
  \centering
  \caption{
    Automatic evaluation of KG metadata compatible and incompatible entity-predicate pairs. 
    \#Pairs refers to the number of overlapping pairs between predictions and test sets, 
    and Precision is their precision scores: 
    KG metadata guidance can help prediction, 
    because compatible groups show higher precision than incompatible groups. 
   }
  \label{tab:other_guidance_precision}
  \begin{tabular}{lrcrc}
    \toprule
    \multirow{2}{*}{} & \multicolumn{2}{c}{DBPedia900K} & \multicolumn{2}{c}{YAGO5M} \\
                 & \#Pairs & Pair Precision & \#Pairs & Pair Precision \\
    \midrule
    Incompatible & 3553456 & 0.0078    & 8216555 & 0.0209 \\
    Compatible   & 424526  & 0.0337    & 139554  & 0.0488 \\
    \bottomrule
  \end{tabular}
\end{table}

\begin{figure}
  \centering
  \includegraphics[width=0.55\linewidth]{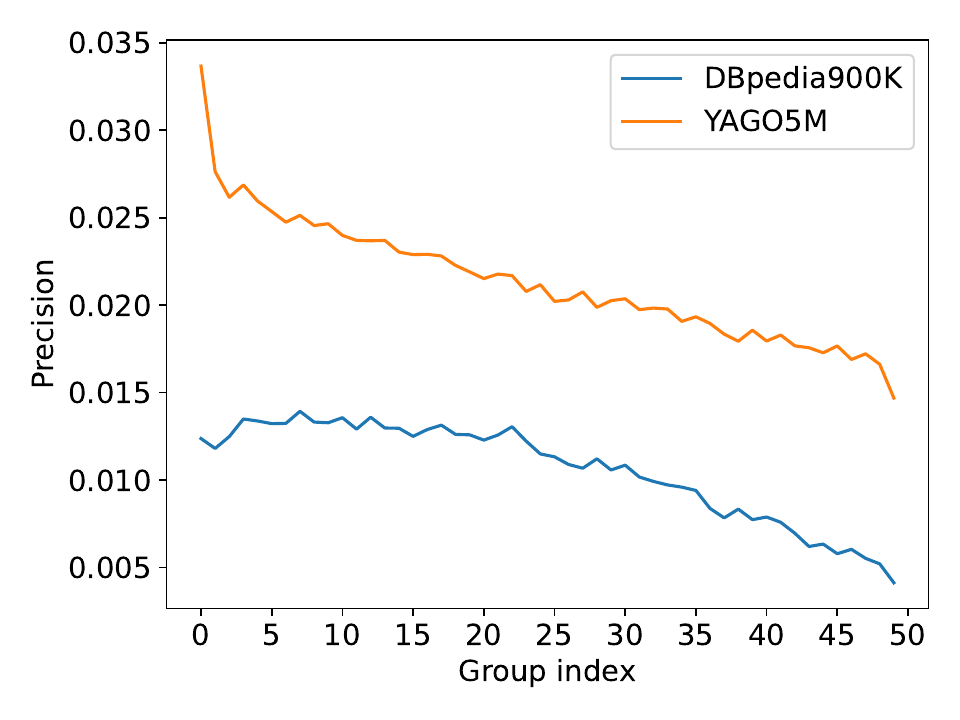}
  \caption{
    Automatic evaluation of embedding score guidance. 
    Y-axis is the precision score of each group, 
    and X-axis shows the indices of the groups sorted by embedding scores, 
    in which the larger is the group index the lower is the embedding score:  
    embedding score guidance can help missing triplet prediction, 
    but worse than query guidance.
    }
  \label{fig:embedding_score_guidance}
\end{figure}

\paragraph{Automatic Evaluation}

We show the automatic evaluation results for KM in Table~\ref{tab:other_guidance_precision}, 
where \#Pairs and Pair Precision are the number of predicted entity-predicate pairs in this group 
(KG metadata compatible and incompatible) and their corresponding precision scores. 
We also show the automatic evaluation results for ES in Figure~\ref{fig:embedding_score_guidance}, 
where x-axis is the group index, and larger group index indicates lower embedding score, 
which indicates lower quality 
(recall that we divide all predicted entity-predicate pairs into 50 bins based on their embedding scores); 
and y-axis is the precision score of this group. 

We make three observations. 
First, the guidance of both KG metadata and embedding score can help prediction.   
This observation is supported by that 
1) in Table~\ref{tab:other_guidance_precision}, 
the precision scores of the compatible groups are $>2$ times higher 
than those of the incompatible groups; 
and 2) in Figure~\ref{fig:embedding_score_guidance}, 
groups with higher embedding scores (i.e. smaller group indices) are of higher precision scores. 
Second, user queries still provide better guidance than 
both KG metedata and embedding scores, shown by that both 
1) the precision scores of the compatible groups in Table~\ref{tab:other_guidance_precision} and 
2) the group of the highest embedding score in Figure~\ref{fig:embedding_score_guidance} (i.e. leftmost)
are outperformed by QG (i.e. Pair Precision in Table~\ref{tab:precision_results}). 
Third, we observe that 
only a small portion of the predicted entity-predicate pairs are compatible with KG metadata. 
Considering that KM works in a post-hoc way 
(i.e., it filters out incompatible ones after predictions are made), 
this result suggests the relatively low efficiency of KM compared with QG. 
The same concern applies to ES if we solely rely on the a few groups with the highest embedding scores.

\paragraph{Human Evaluation}

We also conduct a human evaluation study 
to further compare the impact of these two types of guidance against query guidance. 
Consistent with \S\ref{subsec:human_eval}, for each KG, 
we randomly select 200 entity-predicate pairs from 
both the compatible group in KM and the group of the highest embedding score in ES, 
and annotate their correctness and relevance. 

Table~\ref{tab:human_eval} shows the results. 
We make similar observations as for the automatic evaluation: 
while the guidance through both KG metadata and embedding score achieve improvements over baseline, 
they are outperformed by QG.

\section{Conclusions and Limitations}

To improve the precision and relevance of KGC methods, 
we propose a user-driven approach based on explorative query logs. 
Our approach conceptually works for any type of query language where entities and properties can be defined. 
This includes explicit definition as RDF constructs in SPARQL, 
or implicitly through natural language queries ``make-up for dark skin tone''. 
The latter is particularly interesting for catching up with user-defined trends regarding product attributions.
Because commercial KGs and queries are usually confidential, 
we perform our experiments with two popular general-purpose KGs, DBPedia and YAGO 4, 
and their SPARQL user queries. 
Specifically, we extract entity-predicate pairs from SELECT queries, 
and make predictions from KGE models from them, 
for they are likely to be correct and relevant to users. 
Our results from both automatic and human evaluations show that  
query guidance can significantly improve the correctness and relevance of predicted facts.  

Our approach and its adaptation for open KGs opens up further avenues 
for the combined usage of KGs and query logs. 
In particular, 
future work can explore further aggregation and filtering of queries, 
and harvest more sophisticated structures from complex queries that suggest missing facts.

\bibliographystyle{splncs04}
\bibliography{custom.bib}

\end{document}